\documentclass[12pt,preprint]{aastex}

\usepackage{emulateapj5}

\shorttitle{FIRST CV}
\shortauthors{Bond et al.}

\begin{document}

\title{FIRST J102347.6+003841: The First Radio-Selected Cataclysmic Variable}

\author{Howard E. Bond\altaffilmark{1},
Richard L. White\altaffilmark{1},
Robert H. Becker\altaffilmark{2},
and M. Sean O'Brien\altaffilmark{1}}

\altaffiltext{1}
{Space Telescope Science Institute, 
3700 San Martin Dr.,
Baltimore, MD 21218; 
bond@stsci.edu, rlw@stsci.edu, obrien@stsci.edu}

\altaffiltext{2}
{Department of Physics, 
University of California, 
Davis, CA 95616; and
IGPP/Lawrence Livermore National Laboratory;
bob@igpp.ucllnl.org}

\begin{abstract}

We have identified the 1.4~GHz radio source FIRST J102347.6+003841 (hereafter
FIRST J1023+0038) with a previously unknown 17th-mag Galactic cataclysmic
variable (CV)\null.  The optical spectrum resembles that of a magnetic
(AM~Herculis- or DQ~Herculis-type) CV\null.  Five nights of optical CCD
photometry showed variations on timescales of minutes to hours, along with
rapid flickering.  A re-examination of the FIRST radio survey data reveals that
the radio detection was based on a single 6.6~mJy flare; on two other occasions
the source was below the $\sim$1~mJy survey limit.  Several other magnetic CVs
are known to be variable radio sources, suggesting that FIRST J1023+0038 is a
new member of this class (and the first CV to be discovered on the basis of
radio emission).  However, FIRST J1023+0038 is several optical magnitudes
fainter than the other radio-detected magnetic CVs.  It remains unclear whether
the source simply had a very rare and extraordinarily intense radio flare at
the time of the FIRST observation, or is really an unusually radio-luminous CV;
thus further observations are urged.

\end{abstract}



\section{Introduction}

Most of the known cataclysmic variables (CVs) were originally discovered
because of their dramatic optical variability, exemplified by the outbursts of
dwarf novae and classical novae. More recently, however, ultraviolet and X-ray
sky surveys have led to discoveries of many new CVs, including members of
subclasses that have less spectacular optical light variations (e.g., the
novalike variables).

Follow-up investigations at radio wavelengths have led to detections of a few
well-known, mostly nearby, CVs as radio sources.  In this paper we report what
we believe to be the first case in which the sequence of discovery has been
reversed: we have detected a new CV during a radio survey, with the subsequent
confirmation having been made in the optical band.

\section{The Discovery of FIRST J1023+0038}

The FIRST (``Faint Images of the Radio Sky at Twenty Centimeters'') survey is
aimed at producing a deep radio map of 10,000 square degrees at the north and
south galactic caps (see Becker, White, \& Helfand 1995; White, Becker,
Helfand, \& Gregg 1997).  The survey is being carried out at 1.4~GHz with the
Very Large Array (VLA) of the National Radio Astronomy Observatory (NRAO), and
can be thought of as a radio analog of the optical Palomar Observatory Sky
Survey (POSS)\null. Source positions are provided to better than $1''$
accuracy, and the sensitivity limit is about 1~mJy.

The FIRST Bright Quasar Survey (FBQS; Gregg et al.\ 1996, White et al.\ 2000;
Becker et al.\ 2001) is a follow-up effort aimed at finding large numbers of
QSOs and characterizing their properties. The FBQS candidates are FIRST radio
sources that have a stellar-appearing optical counterpart in the APM catalog,
which is based on Automatic Plate Measuring machine scans of the POSS-I plates
(McMahon \& Irwin 1992; McMahon et al.\ 2002).  The majority of FBQS candidates
are found to be QSOs, BL Lac objects, or other extragalactic sources. Only 8\%
of the candidates turn out to be Galactic stars, the bulk of which are merely
chance coincidences of unrelated, ordinary foreground stars with the radio
positions. Actual radio emission from Galactic stars is exceedingly rare, even
at the 1~mJy limit of the FIRST survey, as discussed in detail by Helfand et
al.\ (1999).

It was thus a surprise when one of the faint FBQS candidates proved to be a
Galactic star with an unusual spectrum, which clearly showed it to be a
previously unknown CV\null. This source, cataloged with a 1.4~GHz radio flux
density of 3.58~mJy, lies at the J2000 coordinates given in the first line of
Table~1. Lying $1\farcs6$ away from the radio source, at the coordinates listed
in the second line of Table~1, is a faint stellar object contained in the APM
catalog.   Also listed in the bottom line of Table~1 are the coordinates and
magnitudes of a nearby comparison star, used in the CCD observations described
in the next section. Figure~1 presents a finding chart for the optical
candidate, with the radio contours superposed.  The radio source is designated
FIRST J102347.6+003841 (but hereafter the name will be shortened to FIRST
J1023+0038).



Figure~2 shows the discovery spectrogram, which was obtained on 2000 May 6 by
Mark Lacy with the Lick Observatory 3-m telescope.  The spectrum is dominated
by a blue continuum, with superposed emission lines of the Balmer series,
\ion{He}{1}, and \ion{He}{2}.  The spectrum is fairly typical of a CV in
quiescence (see, for example, the spectroscopic atlas of Williams 1983). 
However, the lines of \ion{He}{1} and especially of \ion{He}{2} are rather
stronger than in most dwarf novae, and suggest that the object may be more
closely related to the subset of magnetic or AM Herculis-type CVs.  The
spectrum is, in fact, rather similar to that of AM Her itself, as illustrated
in Williams' Figure~13. 


As this paper was being prepared for publication, we realized that the source
lies within the area of the sky covered by the June 2001 Early Data Release of
the Sloan Digital Sky Survey (SDSS; Stoughton et al.\ 2002). Inspection of the
data archive (maintained at STScI) shows that both photometry and a spectrum
are available. Table~2 gives the SDSS astrometry and photometry for both the CV
candidate and the comparison star.  The SDSS position is $1\farcs1$ away from
the radio position.  The SDSS spectrum is quite similar to our Lick spectrum
shown in Figure~2, although the equivalent widths of the emission lines appear
to be slightly lower.  Moreover, the emission lines are incipiently resolved
into double-peaked profiles.  This would suggest that there is an accretion
disk present in the system, and that therefore it may be a DQ~Herculis-type
magnetic CV, rather than a full-fledged AM~Herculis system (in which an
accretion disk would be lacking).

\section{Optical Photometry}

If the optical candidate counterpart of FIRST J1023+0038 is a CV, then it would
be expected to exhibit rapid optical variability.  In order to search for such
variations, H.E.B. and M.S.O. obtained photometric observations on five nights
in 2000 November and December, using the 2.1-m reflector at Kitt Peak National
Observatory (KPNO)\null.  A Tektronix $2048\times2048$ CCD was used, with
$2\times2$ on-chip binning.  In order to maximize the signal, we used a Schott
BG39 glass filter, which yields a broad-band blue bandpass. Only a
$512\times512$ pixel subraster was read out, in order to reduce the read-out
time between exposures but still produce a large enough field to contain both
the target and a nearby comparison star. Even with this CCD setup, however, the
readout time was still 25~s with the KPNO data acquisition system.  Exposure
times of 30-60~s were used in 2000 November, and were shortened to 10~s for all
of the 2000 December observations.  

Table~3 contains a log of the CCD observations. As noted in the table, none of
the nights were photometric, but accurate differential photometry is readily
possible under non-photometric conditions, as has been shown by Grauer \& Bond
(1981), Ciardullo \& Bond (1996), and many others.  We performed standard
aperture photometry on the frames, and reduced the data to differential
magnitudes of the target with respect to the comparison star.  In order to set
an approximate photometric zero point, we adopted the blue magnitude of 16.14
given for the comparison star in the APM catalog.

Figure~3 shows our resulting light curves from the five nights of photometry.
The differential magnitudes are generally accurate to better than 0.01~mag
(i.e., smaller than the size of the plotting points), except on the very cloudy
night of 2000 Dec 24 where the errors were typically several hundredths of a
magnitude.

The object indeed shows variability on both slow and rapid timescales,
confirming its classification as a CV\null. Although generally maintaining a
level around magnitude 17.5, it showed a flare of $\sim$0.8~mag lasting about
20~min (apart from one rapid dip back to the baseline level) on 2000 December
23. On December 24 the target was more active, being initially about 0.5~mag
brighter than on previous nights, declining briefly, and then ramping up by
about 1~mag over about half an hour and remaining at that level for over an
hour (apart from two rapid dips).

The object also shows short-timescale flickering, which gives the light curves
a noisy appearance.  On the first four nights, and on the fifth night when the
star was near its baseline level, the light curves show a curious ``bimodal''
appearance, i.e., the magnitudes tend to cluster near either ``high'' or
``low'' levels but rarely in between.  We suspect that this appearance 
arises from our sparse sampling of a light curve that may have flickering with a
timescale of approximately 10~s; we are thus missing about 2/3 of the points
that we would have obtained with uninterrupted 10-s integrations.

Power spectra were calculated for all of the light curves, but revealed no
significant coherent signals; of course, we were not able to test for signals
with periods shorter than $\sim$70~s, such as are seen in some of the
DQ~Her-type subset of magnetic CVs (Patterson 1994), because of the $\ge$35~s
interval between integrations.

We see no evidence, in our fairly limited data set, for eclipses or other
repetitive phenomena occurring on an orbital timescale.  Determination of the
orbital period of the system will require additional photometry or a
radial-velocity study.

\section{Radio Variability}

A strong confirmation that the optical CV is indeed the counterpart of the
radio source would come from variability of the radio source.

We therefore searched the FIRST database for evidence of such variability. The
FIRST survey is based on data collected by the VLA on a hexagonal grid of field
centers spaced by about $26'$, and an object can thus appear in overlapping
images from several adjacent fields (Becker, White, \& Helfand 1995).  The
FIRST source catalog is constructed by summing the overlapping fields to get
uniform sensitivity on the sky, and does not contain information on source
variability.  However, a sufficiently bright radio source can be detected by
inspecting the individual grid images before they are coadded.  The sensitivity
in each observation is a function of the distance of the source from the field
center.

Table~4 lists details of the three VLA  observations of fields that contained
FIRST J1023+0038. The successive columns give the observation date and time,
followed by the position of the field center.  Col.~5 gives the distance of the
source from the field center, and col.~6 contains the primary beam response of
the VLA at that radius.  (Note, for example, that the response drops to 0.5 of
its value at the field center at a radius of $15'$.) Finally, in cols.~7 and 8,
$F_{\rm obs}$ is the measured flux density (mJy) for the source (or the
$5\sigma$ upper limit in the case of non-detection), and $F_{\rm corr}$ is the
true flux density after correction for the primary beam response.

As Table~4 shows, FIRST J1023+0038 was detected in only one of the three
observations, on 1998 August 10. At that time,  the implied true flux density
of 6.6~mJy was well above the $5\sigma$ upper limits of the two earlier
observations.  The source is therefore definitely variable at 1.4~GHz by a
factor of 3 or more over a period of only a few days.  The source was also not
detected in the 1.4~GHz NRAO VLA Sky Survey, carried out between 1993 and 1996,
with a 50\% completeness limit of 2.5~mJy (Condon et al.\ 1998). Variability at
this level among FIRST radio sources is extremely rare, and makes it virtually
certain that the variable radio source is the counterpart of the optical CV.

\section{Discussion}

The first CV to be detected at radio wavelengths was the prototypical magnetic
CV AM Herculis, whose emission was discovered at the VLA by Chanmugam \& Dulk
(1982).  They detected a flux density of 0.67~mJy at 4.9~GHz, and suggested
that the emission is from mildly relativistic electrons trapped in the
magnetosphere of the white dwarf.  Subsequent observations by Dulk,  Bastian,
\& Chanmugam (1983) led to detection of a strong flare from AM~Her, which
lasted about 10~min and reached a peak of 9.7~mJy at 4.9~GHz; Dulk et al.\ 
were also able to set an upper limit on the quiescent flux density of AM~Her at
1.4~GHz of 0.24~mJy.

The magnetic DQ Her-type system AE Aqr has also been detected as a variable
radio source, with the initial discovery having been made at the VLA by
Bookbinder \& Lamb (1987); they found a flux density at 1.4~GHz varying from 3
to 5 mJy.  Other radio detections of AM Her- and DQ Her-type CVs have been
reported by Pavelin, Spencer, \& Davis (1994) and references therein. By
contrast, non-magnetic CVs in general are not detectable radio sources (Nelson
\& Spencer 1988), strongly suggesting that a highly magnetic white dwarf is a
key element in the production of radio emission.  For further information, the
reader is directed to the summary of radio emission from CVs by Mason,
Fisher, \& Chanmugam (1996).

We thus strongly suspect that FIRST J1023+0038 is a new magnetic CV, either of
the DQ~Her or AM~Her variety. Our detection of a 6.6~mJy flare at 1.4~GHz is 
remarkable because, with a quiescent optical magnitude near 17.5, FIRST
J1023+0038 is so optically inconspicuous. (By comparison, AE~Aqr is an 11th-mag
object, and AM Her lies generally around 12th-13th mag, although with
occasional drops to ``low'' states below 15th~mag). It remains to be seen
whether the flare that led to our serendipitous discovery of FIRST J1023+0038
was simply a very rare and unusually energetic one, or whether this object is
indeed much more radio luminous than the typical magnetic CVs.

We urge follow-up observations at radio, optical, and X-ray wavelengths. 
Optical spectroscopy should reveal the orbital period, and might provide direct
evidence for a strong magnetic field.  Polarimetry would determine whether the
object is a highly magnetic AM~Her system.  A search of archival plate
collections, and future long-term photometric monitoring, might reveal either
dwarf-nova outbursts, or AM~Her-like low states, although the object is rather
faint.

\acknowledgments

H.E.B. acknowledges the inspiration of his friend and colleague G.~Chanmugam. We
thank Mark Lacy for obtaining the Lick spectrogram.
R.H.B. acknowledges support from NRAO, NSF (grant 
AST-00-98355), and the Institute of Geophysics and Planetary
Physics (operated under the auspices of the U.S. Department of Energy by
Lawrence Livermore National Laboratory under contract
No.\ W-7405-Eng-48).
Kitt Peak National Observatory is operated by AURA Inc., under contract with
the National Science Foundation.  The Digitized Sky Surveys were produced at
the Space Telescope Science Institute under U.S. Government grant NAG W-2166.




\clearpage

\begin{deluxetable}{lcccc}
\tablewidth{0pt}
\tablecaption{Radio Position and APM Astrometry and Magnitudes}
\tablecolumns{2}
\tablehead{
\colhead{Object} & \colhead{$\alpha_{2000}$} &
\colhead{$\delta_{2000}$}  & \colhead{$m_{\rm blue}$}  & 
\colhead{$m_{\rm red}$}   
}
\startdata
FIRST radio position & 10:23:47.621 & +00:38:41.60 & $\dots$ & $\dots$ \\
Optical candidate    & 10:23:47.713 & +00:38:42.40 & 18.05   & 16.98   \\
Comparison star      & 10:23:43.340 & +00:38:19.20 & 16.14   & 14.88   \\
\enddata
\end{deluxetable}


\begin{deluxetable}{lccclllll}
\tablewidth{0pt}
\tablecaption{SDSS Astrometry and Photometry}
\tablecolumns{2}
\tablehead{
\colhead{Object} & \colhead{SDSS Object ID} & \colhead{$\alpha_{2000}$} &
\colhead{$\delta_{2000}$}  & \colhead{$u'$}  & 
\colhead{$g'$} &  
\colhead{$r'$}  & 
\colhead{$i'$}   &
\colhead{$z'$}   
}
\startdata
Optical candidate & 2255048168702071 & 10:23:47.683 & +00:38:40.99 &
19.418 & 17.794 & 17.206 & 17.053 & 16.999 \\
Comparison star   & 2255048168701986 & 10:23:43.308 & +00:38:19.14 &
16.281 & 14.906 & 15.513 & 14.226 & 14.168 \\
\enddata
\end{deluxetable}

\begin{deluxetable}{lccccl}
\tablewidth{0pt}
\tablecaption{CCD Observing Log, KPNO 2.1-m Telescope}
\tablecolumns{2}
\tablehead{
\colhead{UT Date} & \colhead{Observer} & \colhead{UTC} &
\colhead{Run Length }  & \colhead{Integration}  & 
\colhead{Sky}  \\
\colhead{} & \colhead{   } & \colhead{(start)} &
\colhead{(hr)}  & \colhead{Time (s)}  & 
\colhead{} 
}
\startdata
2000 Nov 21 & M.S.O. & 10:57 &	   	 1.0  &   60, 30 & cirrus \\
2000 Nov 22 & M.S.O. & 11:35 &	   	 1.3  &   30	 & thin haze \\
2000 Dec 22 & H.E.B. & 12:28 &	   	 1.2  &   10	 & thin cirrus \\
2000 Dec 23 & H.E.B. & 10:15 &	   	 3.5  &   10	 & haze \\
2000 Dec 24 & H.E.B. & \phantom{1}9:28 & 3.9  &   10     & thick haze \\
\enddata
\end{deluxetable}

\begin{deluxetable}{lccccccc}
\tablewidth{0pt}
\tablecaption{VLA 1.4 GHz Observations of FIRST J1023+0038}
\tablecolumns{2}
\tablehead{
\colhead{Date} & \colhead{UTC} & \colhead{$\alpha_{2000}$} &
\colhead{$\delta_{2000}$}  & \colhead{Radius}  & 
\colhead{Response} & \colhead{$F_{\rm obs}$} &  
\colhead{$F_{\rm corr}$} \\
\colhead{    } & \colhead{   } & \colhead{} &
\colhead{}  & \colhead{(arcmin)}  & 
\colhead{Function} & \colhead{(mJy)} &  
\colhead{(mJy)} 
}
\startdata
1998 Aug  3 & 20:33 & 10:24.0 & +00:52 &  13.7 &  0.57 & \llap{$<$}1.0 & 
  \llap{$<$}1.8 \\
1998 Aug  8 & 21:08 & 10:22.5 & +00:39 &  19.4 &  0.30 & \llap{$<$}1.0 & 
  \llap{$<$}3.4 \\
1998 Aug 10 & 21:24 & 10:24.0 & +00:26 &  13.1 &  0.60 &  3.9\rlap{1}  &   
  6.5\rlap{6} \\
\enddata
\end{deluxetable}

\clearpage

\begin{figure}
\begin{center}
\includegraphics[width=\hsize]{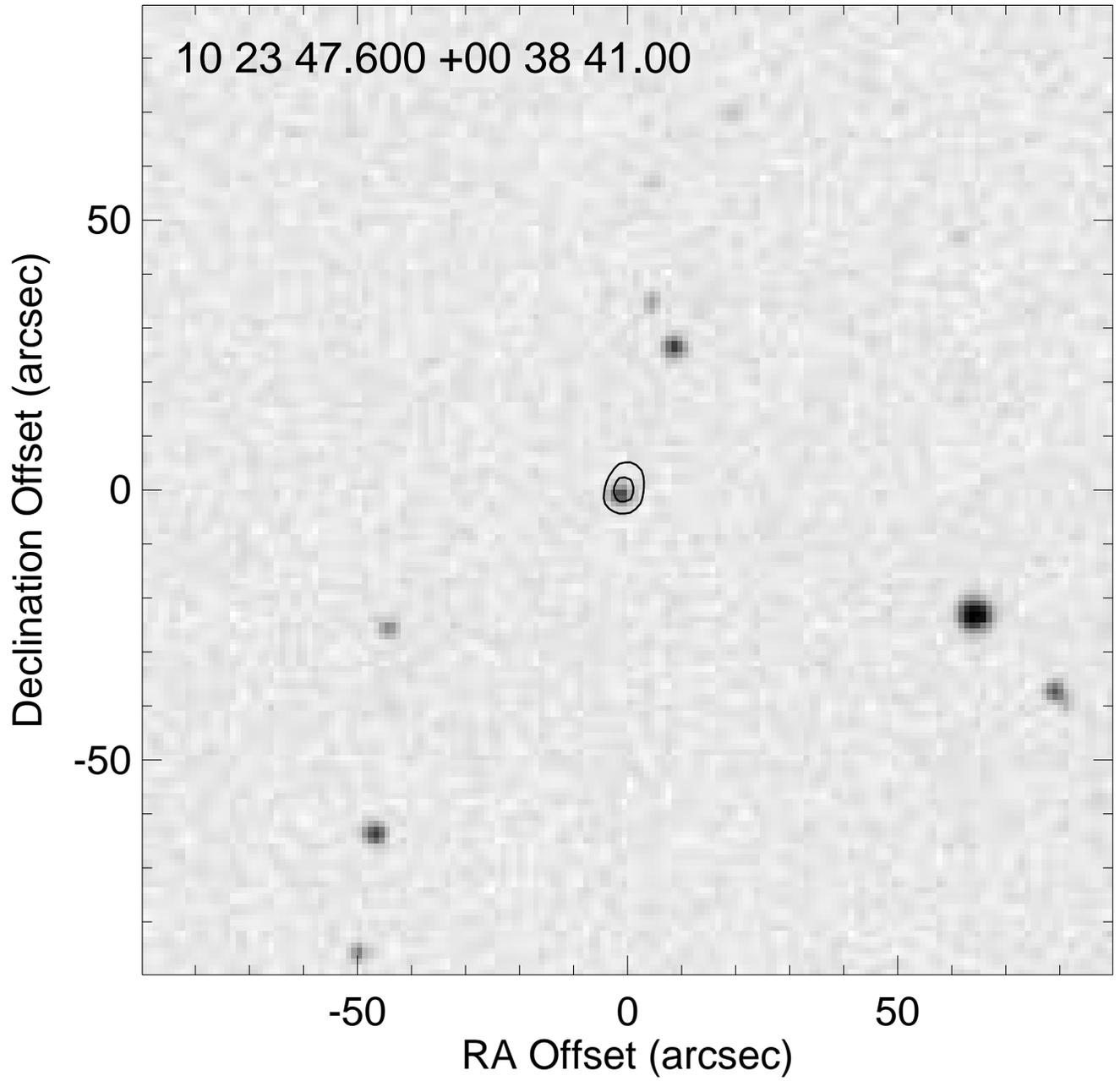}
\end{center}
\figcaption{Finding chart for optical counterpart of FIRST J1023+0038, prepared
from the second-epoch red Digitized Sky Survey POSS scans.
Superposed are 0.8 and 2.0~mJy 1.4~GHz radio contours from the FIRST survey.
North is at the top and east on the left.}
\end{figure}
\clearpage

\begin{figure}
\begin{center}
\includegraphics[width=\hsize]{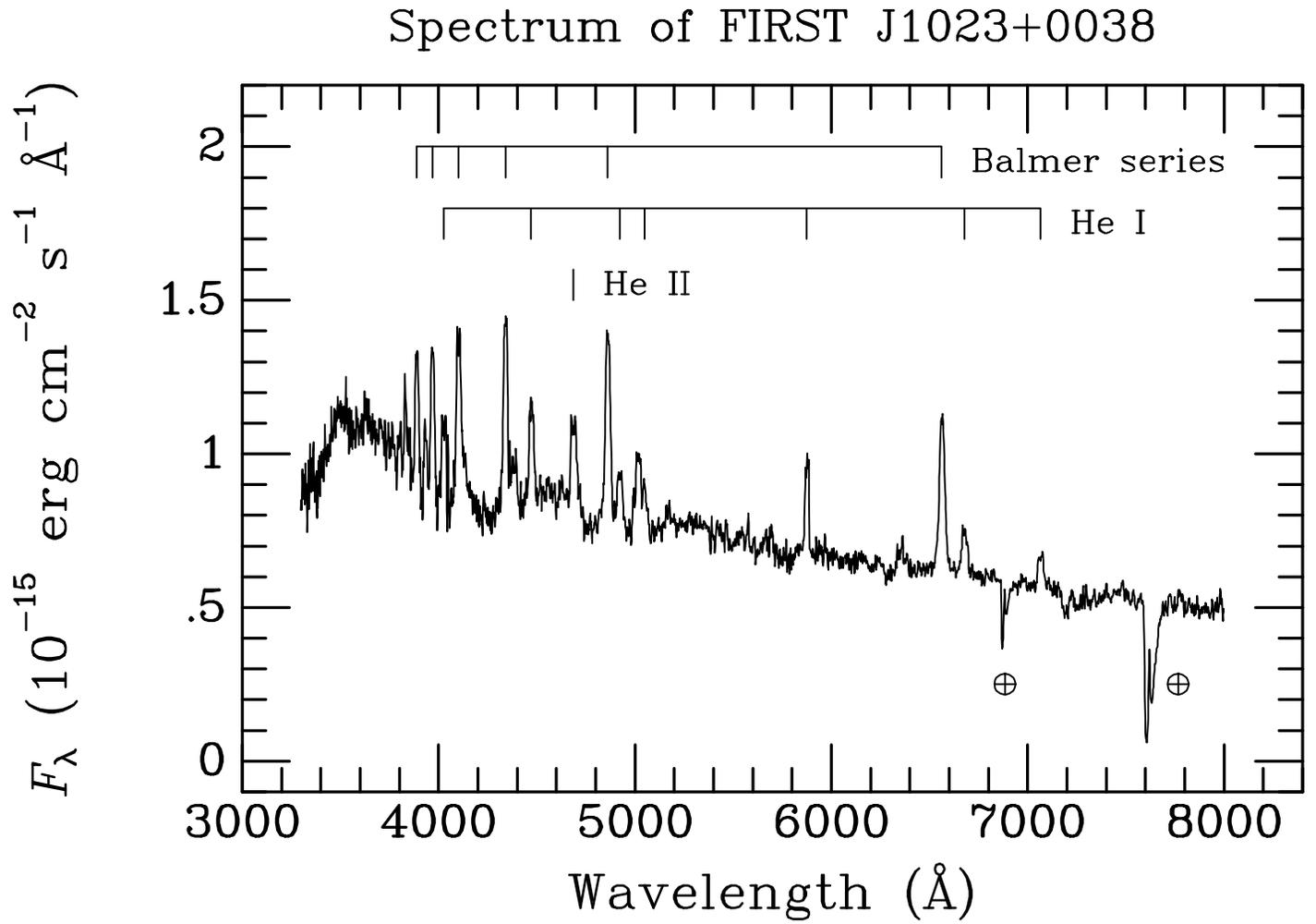}
\end{center}
\figcaption{Lick 3-m spectrum of FIRST J1023+0038. Emission lines of the Balmer
series and of \ion{He}{1} and \ion{He}{2} are marked, as are two terrestrial
atmospheric absorption bands. The spectrum is typical of a cataclysmic variable
in quiescence.}
\end{figure}
\clearpage

\begin{figure}
\begin{center}
\includegraphics[width=\hsize]{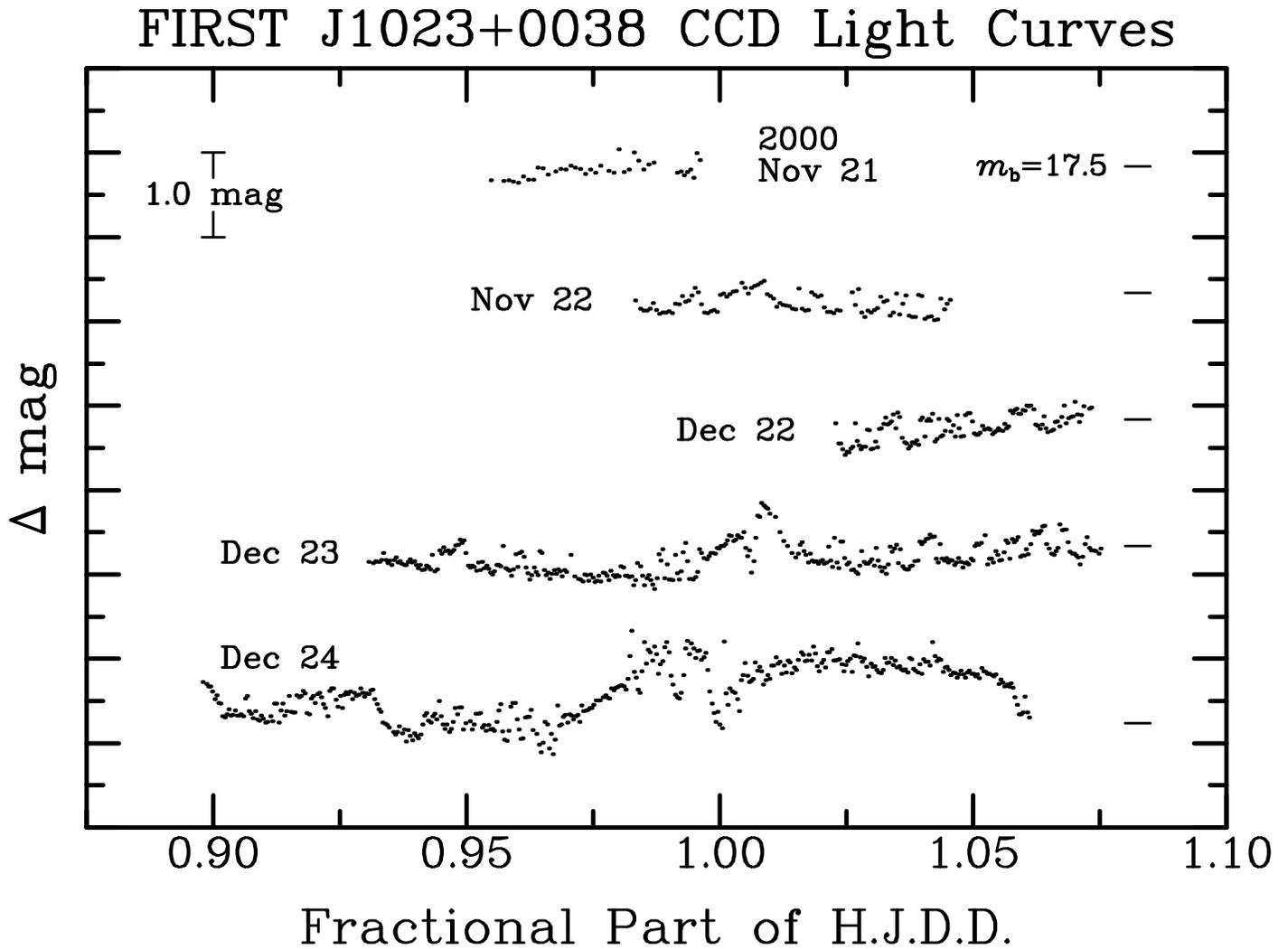}
\end{center}
\figcaption{CCD broad-band blue light curves of FIRST J1023+0038, obtained with
the KPNO 2.1-m telescope and a BG39 filter on five nights in 2000
November-December. UT dates are indicated.  The time axis is in days, with tick
marks at intervals of 0.025~day = 36~min, and the scale bar at the upper left
shows a 1~mag brightness interval.  Differential magnitudes with respect
to a nearby comparison star are plotted, with the approximate 17.5 mag 
zero-point levels for each night indicated by ticks down the right-hand edge. 
Photometric errors are generally smaller than the plotted points.}
\end{figure}
\clearpage

\end{document}